\documentclass[
    ,final            
  ]
  {aipproc}

\layoutstyle{6x9}

\def\d{\mathrm{d}}
\def\ket#1{|#1\rangle}
\def\bra#1{\langle#1|}

\def\psir{\psi}
\def\Hr{{\widehat H}}
\def\qr{{\widehat q}}
\def\zg{z}
\def\Mg{\mathbf{M}}
\def\Mgt{{\mathbf{\widetilde M}}_t}
\def\lg{\mathbf{[}}
\def\rg{\mathbf{]}}
\def\chir{{\widehat\chi}}
\def\Psir{\Psi}
\def\Jr{{\widehat J}}
\def\Jg{J}
\def\Ar{{\widehat A}}
\def\Ag{A}
\def\Agm{A^{\scriptscriptstyle-}}
\def\Agp{A^{\scriptscriptstyle+}}
\def\Cg{C}
\def\Qr{{\widehat Q}}

\begin{document}

\title{Continuous Wave Function Collapse in Quantum-Electrodynamics?}

\classification{03.65.Ta,03.65.Pm03.65.Yz,12.20.-m}
\keywords      {wave function collapse, stochastic Schr\"odinger equation, Lorentz invariance,
                quantum-electrodynamics}

\author{Lajos Di\'osi}{
  address={Research Institute for Particle and Nuclear Physics\\
           H-1525 Budapest 114, P.O.Box 49, Hungary}
}

\begin{abstract}
Time-continuous wavefunction collapse mechanisms  \emph{not}  restricted to
markovian approximation have been found only a few years ago, and have
left many issues open. The results apply formally to the standard relativistic
quantum-electrodynamics. I present a generalized Schr\"odinger equation
driven by a certain complex stochastic field. The equation reproduces the
\emph{exact}  dynamics of the interacting fermions in QED. The state of the
fermions appears to collapse continuously, due to their interaction with
the photonic degrees of freedom. Even the formal study is instructive
for the foundations of quantum mechanics and of field theory as well.
\end{abstract}

\maketitle


\section{Introduction}
The first study of what we can call continuous wave function collapse was Mott's analysis \cite{Mot29}
of the 
particle's track in a cloud chamber. The advanced models \cite{DCM89}-\cite{WM93}
of real continuous collapse have become part of the quantum optics theoretical toolbox. 
Speculations on fictitious or formal continuous wave function collapse started with \cite{BB66} by
Bohm and Bub and have been discussed in numerous works \cite{Kar66}-\cite{WD01} of various motivations. 
As for the mathematical formalisms of both
real and fictitious collapses, the triumphing one turned out to be the 
markovian Stochastic(ally modified) Schr\"odinger Equation (SSE) \cite{Gis84},\cite{Dio88},\cite{Bel88}.
The Lorentz invariant SSE is a hard nut, see e.g. \cite{AB01} and references therein, as well as
\cite{Rim03},\cite{Tum06}. In my opinion, a sensible Lorentz invariant SSE should relax the markovian 
approximation first. After my early attempts
\cite{Dio90}-\cite{Dio96}, Strunz published an important result \cite{Str96} to be followed soon by many 
others \cite{DS97}-\cite{GW03}. 
I am going to recall the standard markovian and non-markovian SSE, and
I shall present and discuss a SSE in the explicit Lorentz invariant context of QED. 

\section{Real or Fictitious Continuous Collapse}
Classicality emerges from Quantum via real or fictitious, often
time-continuous, measurement (detection, observation, monitoring, e.t.c.)
of the wavefunction $\psir$.
By the real continuous collapse we mean, e.g., the detection of a particle track in a cloud chamber, 
the photon-counter detection of atomic decay, or the homodyne detection of quantum-optical oscillators. 
The fictitious continuous collapse means the various theories of spontaneous (universal, intrinsic, primary, e.t.c.) 
collapse (localization, reduction, e.t.c.). The items in the parentheses indicate the multitude of close synonyms.    
To date, the mathematics is the same for both real and fictitious classes! 
We know almost everything about the mathematical and physical structures if markovian approximation applies.
We know much less beyond that approximation. So, let us see
what \emph{equation} describes the wave function under time-continuous collapse?

\section{The Markovian Stochastic Schr\"odinger Equation}
The prototype of the SSE has the following structure:
\begin{eqnarray}
\label{MSSE}
\frac{\d\psir(t,\zg)}{\d t}=&-i\Hr\psir(t,\zg)~~~~         &\mbox{hermitian Hamiltonian}\nonumber\\  
                            &-i\qr\zg\psir(t,\zg)~~~~      &\mbox{non-hermitian noisy Hamiltonian}\nonumber\\
                            &-\frac{1}{2}\gamma\qr^2\psir(t,\zg)~~~~        
                                                           &\mbox{non-hermitian dissipative Hamiltonian},
\end{eqnarray}
where $\zg$ is a complex Gaussian hermitian white-noise:
\begin{equation}
\label{MSSEnoise}
\Mg\lg\zg^\star(t)\zg(s)\rg=\gamma\delta(t-s).
\end{equation}
Throughout this work, $\Mg\lg\dots\rg$ stands for the stochastic average.
The eq.(\ref{MSSE}) is not norm-preserving. We define the physical state by $\psir/\Vert\psir\Vert$ while its 
statistical weight must be multiplied by $\Vert\psir\Vert^2$:
\begin{eqnarray}
\label{MSSEstate}
\psir(t,\zg)&\longrightarrow&\frac{\psir(t,\zg)}{\Vert\psir(t,\zg)\Vert}\equiv\ket{t,\zg},\\
\label{MSSEGirs}
\Mg\lg\dots\rg&\longrightarrow&\Mg\lg\:\Vert\psir(t,\zg)\Vert^2\dots\rg\equiv\Mgt\lg\dots\rg.
\end{eqnarray}
In our case, the state $\ket{t,\zg}$ and the noise $\zg(t)$ play the roles of the Quantum and the
Classical, respectively. Their ``mutual influence'' is described
by the eqs.(\ref{MSSE}-\ref{MSSEGirs})\footnote{Equivalently, there exists a pair of closed non-linear equations
for $\ket{t,\zg}$ and for the recorded value of $\qr$,
cf.~\cite{Dio88},\cite{Bel88}, or \cite{WD01}.}.

The markovian SSE describes perfectly the time-continuous collapse of the wavefunction in the given 
observable(s) $\qr$. The state $\ket{t,\zg}$ depends on $\{\:\zg(s);s\leq t\:\}$ causally,
i.e., the state at $t$ depends on the values of the noise at times $s\leq t$. 
The individual solutions $\ket{t,\zg}$ can, in principle, be realized by time-continuous monitoring 
of $\qr$. Then $\zg(t)$ becomes the classical record explicitly related to the monitored value of $\qr$.

Our key-problems will be: causality, realizability, and Lorentz invariance. 
So far, for the markovian SSE, both causality and realizability hold
but Lorentz invariance is missing.
To make a progress, we need to relax the markovian approximation.

\section{The non-Markovian Stochastic Schr\"odinger Equation}
The key to the non-markovian SSE is that the driving noise is a colored noise:
\begin{equation}\label{NMSSEnoise}
\Mg\lg\zg^\star(t)\zg(s)\rg=\alpha(t-s).
\end{equation}
The corresponding SSE \cite{DS97} contains a memory-term: 
\begin{equation}\label{NMSSE}
\frac{\d\psir(t,\zg)}{\d t}=-i\Hr\psir(t,\zg) -i\qr\zg\psir(t,\zg)
                            +i\qr\int_0^t\alpha(t-s)\frac{\delta\psir(t,\zg)}{\delta\zg(s)}\d s.
\end{equation}
This linear SSE is not norm-preserving. We define the physical state by $\psir/\Vert\psir\Vert$ while its statistical
weight must be multiplied by $\Vert\psir\Vert^2$ exactly the same way as in eqs.(\ref{MSSEstate},\ref{MSSEGirs}) 
for the markovian SSE\footnote{There exists a non-linear non-markovian SSE for $\ket{t,\zg}$ \cite{SDG99}.
There is no equation for anything like the recorded value of $\qr$. To date, there has been no way to define
a classical record.}.

The non-markovian SSE describes the  \emph{tendency}  of time-continuous collapse of the wavefunction in the given
observable(s) $\qr$. The state $\ket{t,\zg}$ depends on $\{\:\zg(s);s\leq t\:\}$ 
causally. The individual solutions $\ket{t,\zg}$ can  \emph{not}  be realized by any known way of
monitoring \cite{GW03}. The non-markovian SSE corresponds mathematically to the influence of a
real or fictitious oscillatory reservoir whose instantaneous Husimi function is sampled stochastically, cf. \cite{DGS00}.
Disappointedly, $\zg(t)$ can  \emph{not}  be  interpreted as a classical record. It only 
corresponds to fictitious paths in the parameter space of the reservoir's coherent states. 

The status of our key problems for the non-markovian SSE is the following. Causality holds, but realizability and  
Lorentz invariance may not hold at all. Can we enforce Lorentz invariance at least? 

\section{Case study: quantum-electrodynamics}
We choose the well-known quantum theory of electromagnetic interaction as the framework to study the
possible form of a Lorentz invariant SSE --- without any guarantee that it exists. At least, we shall try to export
the Lorentz invariance from QED to SSE.

Let $x=(x_0,\vec x)$ denote the four-vector of space-time coordinates. The vector-field
$\Ar(x)$ stands for the quantized electromagnetic four-potential, and the Dirac spinor field  
$\chir(x)$ stands for the quantized electron-positron field. Then the fermionic current is defined by
$\Jr(x)=e\overline{\chir}(x)\gamma\chir(x)$.
Later we shall need the electromagnetic correlation 
$D(x)=i\bra{\mathrm{e.m. vac}}\Ar(x)\Ar(0)\ket{\mathrm{e.m. vac}}$ as well. 
The Schr\"odinger equation in interaction picture takes this form:
\begin{equation}
\label{QEDint}
\frac{\d\Psir(t)}{\d t}=-i\int_{x_0=t}\!\!\!\!\!\!\!\!\d{\vec x}\:\Jr(x)\Ar(x)\:\Psir(t). 
\end{equation}
As usual in QED, we suppose the uncorrelated initial state 
$\Psir(-\infty)=\psir(-\infty)\otimes\ket{\mathrm{e.m. vac}}$ where $\psir(-\infty)$ is the initial state
of the electrons and positrons before the interaction is switched on. We seek the SSE for the 
electron-positron wavefunction $\psir(t)$ continuously localized by the electromagnetic field which plays
the role of the ``environment'' or the ``reservoir''. It can be shown that the SSE is driven by the
negative-frequency part $\Agm(x)$ of the e.m. ``vacuum-field'' $\Agp+\Agm=\Ag$, 
satisfying:
\begin{equation}
\label{QEDSSEnoise}
\Mg\lg\Agm(x)\Agp(y)\rg=\bra{\mathrm{e.m. vac}} \Ar(x)\Ar(0)\ket{\mathrm{e.m. vac}}=-iD(x-y).
\end{equation}
The SSE\footnote{This equation follows from the results of \cite{DS97}-\cite{DGS00},
where a closed non-markovian SSE for the normalized state $\ket{t,\Ag}$ is also given.} contains a memory-term: 
\begin{equation}
\label{QEDSSE}
\frac{\d\psir(t,\Agm)}{\d t}=-i\!\!\int_{x_0=t}\!\!\!\!\!\!\!\!\d{\vec x}\Jr(x)\Agm(x)\:\psir(t,\Agm)
                               -\int_{x_0=t}\!\!\!\!\!\!\!\!\d{\vec x}\int_{y_0\leq t}\!\!\!\!\!\!\!\!\d y\:\Jr(x)D(x-y)
\frac{\delta\psir(t,\Agm)}{\delta\Agm(y)}.
\end{equation}
This linear non-markovian SSE is not norm-preserving. We define the physical state by $\psir/\Vert\psir\Vert$ 
while its statistical
weight must be multiplied by $\Vert\psir\Vert^2$, cf. eqs.(\ref{MSSEstate},\ref{MSSEGirs}): 
\begin{eqnarray}
\label{QEDSSEstate}
\psir(t,\Agm)&\longrightarrow&\frac{\psir(t,\Agm)}{\Vert\psir(t,\Agm)\Vert}\equiv\ket{t,\Ag},\\
\label{QEDSSEGirs}
\Mg\lg\dots\rg&\longrightarrow&\Mg\lg\:\Vert\psir(t,\Agm)\Vert^2\dots\rg\equiv\Mgt\lg\dots\rg.
\end{eqnarray}

Similarly to the markovian SSE, the state $\ket{t,\Ag}$ and the random complex field $\Agm(x)$ play the role
of the Quantum and the Classical, respectively. 
Their ``mutual influence'' is described by the eqs.(\ref{QEDSSEnoise}-\ref{QEDSSEGirs})
which are \emph{formally} Lorentz invariant.
The solutions of the ``relativistic'' SSE (\ref{QEDSSE}), when averaged over $\Agm$, describe the exact 
QED fermionic reduced state:
\begin{equation}\label{QEDred}
\Mg\lg\psir(t,\Agm)\psir^\dagger(t,\Agp)\rg=\mathrm{tr_{e.m.}}[\Psir(t)\Psir^\dagger(t)].
\end{equation}

The ``relativistic'' SSE describes the  \emph{tendency}  of time-continuous collapse of the fermionic 
wavefunction in the current $\Jr$ although the collapse happens in (certain) Fourier components instead of
the local values $\Jr(x)$. The state $\ket{t,\Ag}$ depends on the classical
field $\{\:\Ag(x);x_0\leq t\:\}$ causally. 
The individual solutions $\ket{t,\Ag}$ can  \emph{not}  be realized by any known way of monitoring. 
Therefore the classical field $\Ag$ can  
\emph{not}  be interpreted as a classical record\footnote{The complex random field $\Agm$ carries information 
on the quantized e.m. field $\Ar(x)$, the details are still to be investigated.}.

\section{Lorentz invariance?}
We can express the solution of the ``relativistic'' SSE (\ref{QEDSSE}), emerging from the initial state
$\psir(-\infty)$: 
\begin{equation}\label{QEDSSEsol}
\psir(t,\Agm)=\mathrm{T}
\exp\left\{-i\!\!\int_{x_0\leq t}\!\!\!\!\!\!\!\!\d x\Jr(x)\Agm(x)
-\int\!\!\int_{y_0\leq x_0\leq t}\!\!\!\!\!\!\!\!\!\!\!\!\!\!\!\!\d x\d y\:\Jr(x)D(x-y)\Jr(y)
\right\}\psir(-\infty),
\end{equation}
where $\mathrm{T}$ stands for time-ordering of the current operators $\Jr(x)$.
Consider the expectation value of the local e.m. current at some $t$:
\begin{equation}\label{QEDJexp}
\Jg(t,\vec x,\Ag)=\frac{\psir^\dagger(t,\Agp)\Jr(t,\vec x)\psir(t,\Agm)}{\psir^\dagger(t,\Agp)\psir(t,\Agm)}.
\end{equation}
This local current  $\Jg(x,\Ag)$ depends on $\Ag(y)$ for $y_0\leq x_0$ which is causal in the given frame while it
may violate causality in other Lorentz frames. To assure Lorentz invariant causality, the local current
$\Jg(x,\Ag)$ must not depend on $\Ag$ outside the backward light-cone of $x$. Since this is not the case,
our ``relativistic'' SSE can not be causal at all. 

We can see that, despite our efforts, the status of the key-problems for the ``relativistic'' SSE
is disappointing: causality, realizability, and Lorentz invariance have been all lost. However, the
``relativistic'' SSE is the prototype of a Lorentz-invariant-looking closed SSE and we must study it
if we wish to know why and where exactly Lorentz 
invariance has gone\footnote{Lorentz invariance would already be lost in a single 
von Neumann-L\"uders collapse. Nevertheless, the status of Lorentz invariance in continuous collapse
is a theoretical challenge.}.

\section{Outlook}
``Classicality emerges from Quantum via real or fictitious, often
time-continuous, measurement (detection, observation, monitoring, e.t.c.)
of the wavefunction $\psir$.'' This has been our universal motivation to investigate the corresponding mathematical
models. All markovian SSE's turn out to be mathematically equivalent with standard (though sophisticated) 
quantum measurements \cite{WD01}. The non-markovian SSE's are equivalent with certain quantum reservoir dynamics, i.e., 
with their formal stochastic decompositions (unravelings) \cite{DS97}. We have inspected in the previous section
that the causality and Lorentz invariance (as well as the realizability) remain problematic even when we start from a true Lorentz 
invariant dynamics.

Let us ask the following question. Can we construct more general models which would liberate
us from the mathematical constraints of the standard quantum theory? The minimalist's answer would be this. 
We should replace the concept ``Emergence of Classicality from Quantum'' by the concept
``Coexistence of Classical and Quantum''. 
The classical entities\footnote{According to certain alternative concepts 
\cite{Dio88},\cite{Tum06},\cite{Dio91},\cite{DH05} the whole physics might be represented 
by classical entities so that we may not care if the wave function violates Lorentz invariance and causality.}   
are certain classical fields $\Cg(x)$ and 
the quantum entities are certain quantum fields $\Qr(x)$. We seek a causal and Lorentz invariant coexistence
including their ``mutual influence'' on each other. The loophole is that, unlike in the quantum theory, the 
``mutual influence'' is not necessarily a dynamical or a measurement-like mechanism. 
In our longstanding struggles with the problem of Classical vs. Quantum,  the main issue to overcome has always been 
the painful lack of a consistent model that ``couples'' the coexisting classical and quantum entities. 
Aren't quantum dynamics and measurement too restrictive? Are there any other consistent mechanisms?


\begin{theacknowledgments}
I thank the organizers of the conferences in Triest and in Mali Losinj
for the kind invitation and the financial support. My research is supported
by the Hungarian OTKA under Grant No. 49384.
\end{theacknowledgments}

\end{document}